\documentclass[11pt]{article}

\usepackage{hyperref}
\usepackage{amsmath,amsfonts,amssymb}
\hypersetup{dvips,dvipdfm,colorlinks=true,urlcolor=magenta,filecolor=magenta,linktoc=page,citecolor=red,linkcolor=blue,bookmarks=true}
\def\mytitle#1{\setcounter{equation}{0}
\setcounter{footnote}{0}
\begin{flushleft}\Large\textbf{#1}\end{flushleft}
}
\def\myname#1{\leftline{{\large #1}}\vspace{-0.1cm}}
\def\myplace#1#2{\small\begin{flushleft}\textit{#1}\vspace{-0.2cm}
\texttt{#2}\end{flushleft}}

\def\myclassification#1{\small\noindent
~~~~~Keywords : naked singularity; black hole; trapped surface; geodesic
	
PACS no : 04.20.-q, 04.30.-w, 04.40.Nr#1}
\usepackage{graphicx}
\usepackage{setspace}
\begin{document}

\mytitle{Collapse Geometry in Inhomogeneous FRW model}

 \myname{Sanjukta Chakraborty\footnote{sanjuktachakraborty77@gmail.com}}
\vskip0.2cm \myname{Akash Bose\footnote{bose.akash13@gmail.com}}
\vskip0.2cm \myname{Subenoy Chakraborty\footnote{schakraborty.math@gmail.com}}
\myplace{$^1$ Department of Mathematics, Acharya Jagadish Chandra Bose College, Kolkata, India.}{}
\myplace{$^{2,3}$ Department of Mathematics, Jadavpur University, Kolkata-700 032, India.}{}
\begin{abstract}
Collapsing process is studied in special type of inhomogeneous spherically symmetric space-time model (known as IFRW model), having no time-like Killing vector field. The matter field for collapse dynamics is considered to be perfect fluid with anisotropic pressure. The main issue of the present investigation is to examine whether the end state of the collapse to be a naked singularity or a black hole. Finally, null geodesics is studied near the singularity.

\end{abstract}
\myclassification{}
{\setstretch{1.3}
	\vspace{-.2cm}
\section{Introduction}
\paragraph{ }As Einstein field equations are second order quasilinear coupled partial differential equations so one has to impose symmetries\cite{Chakraborty05} on space-time to have solutions of these non-linear equations. In cosmology, ``The cosmological principle" is such an assumption on space-time and one has FLRW model. However, spatial homogeneity is one of the reasonable assumptions (in an average sense) for cosmological phenomena over galactic scale but in small scale inhomogeneous solutions may be useful. The inhomogeneous FRW (IFRW) model \cite{Bhandari19}-\cite{Cataldo8} is an example of such space-time model. Geometrically, this space-time has invariant family of spherical hypersurfaces and it has no time-like killing vector field.

The main issue of studying gravitational collapse is to support or disprove the cosmic censorship conjecture (CCC). So far there are lot of works on gravitational collapse with Lemaitre-Tolman-Bondi (LTB) spherically symmetric space-time and quasispherical Szekeres space-time\cite{Chakraborty05}. A general conclusion from these studies is that a central curvature singularity forms but its local or global visibility depends on the initial data. 

On the other hand, as each $t =$ constant space-like hypersurfaces for the present IFRW space-time model are not strictly spherical so there is ambiguity in the formation of horizon. One may use hoop conjecture (by Throne) \cite{Throne72} to characterize the formation of horizon but only few works \cite{Shapiro91}-\cite{Iguchi00} have been done to confirm or refute the conjecture. In the present work, a detailed study of collapse dynamics has been done for IFRW model to characterize the end state of collapse. The plan of the paper is as follows. Section 2 describes the collapsing process in IFRW space time model. Collapse dynamics has been studied and trapped surface formation has been examined in section 3. Geodesics near the singularity are studied in section 4. Finally the paper ends with a shorty summary and remarks in section 5.
\vspace{-.2cm}
\section{The IFRW Space-time model and collapsing process.}

\paragraph{  }The line element for the inhomogeneous Friedmann-Robertson- Walker (IFRW) space-time is given by 
\begin{equation}\label{eq1}
d{s^2}=-d{t^2} +{a^2(t)}\left[\frac{d{r^2}}{1-b(r)} +{r^2}(d{\theta^2} +{sin^2\theta}d{\phi^2})\right]
\end{equation}

where $a(t)$ is the scale factor and $b(r)$ ($\neq \lambda r^{2},$ $\lambda,$ a constant, $b(r)=\lambda r^{2}$ gives the FLRW model) is an arbitrary functions of r alone. 

For perfect fluid having both radial and transverse stresses, the energy-momentum tensor has the following structure 
\begin{equation}\label{eq2}
T_{\mu}^{\nu}=diag (\rho, -p_r, -p_t, -p_t)
\end{equation}

 So the explicit form of the Einstein field equations are
\begin{eqnarray}
3\frac{\dot{a}^{2}}{a^{2}}+\frac{r{b^{'}(r)}+b(r)}{r^{2}a^{2}}&=&8\pi G \rho\label{eq3}\\
2\frac{\ddot{a}}{a}+\frac{\dot{a}^{2}}{a^{2}}+\frac{b(r)}{r^{2}a^{2}}&=&-8\pi G p_{r}\label{eq4}\\
2\frac{\ddot{a}}{a}+\frac{\dot{a}^{2}}{a^{2}}+\frac{b^{'}(r)}{2r a^{2}}&=&-8\pi G p_{t}\label{eq5}
\end{eqnarray}

where an overdot and dash stands for partial differentiation with respect to `$t$' and `$r$' respectively. The energy momentum conservation relation : ${T^{\nu}_{\mu}}_{;\nu}=0$ gives 
\begin{equation}\label{eq6}
\dot{\rho}+(\rho+p_{r})\frac{\dot{a}}{a}+2(\rho+p_{t})\frac{\dot{a}}{a}=0
\end{equation}

and\vspace{-.2cm}
\begin{equation}\label{eq7}
\acute{p_{r}}+\frac{2}{r}(p_{r}-p_{t})=0
\end{equation}

Now, introducing 
\begin{equation}\label{eq8}
\Upsilon(r,t)=R(\dot{R}^{2}+b(r))
\end{equation}

where $R=ar$ is the area radius, the above field equations can be written in compact form as
\begin{eqnarray}
\rho(r,t)&=&\frac{\Upsilon^{'}(r,t)}{R^{2}R^{'}}\label{eq9}\\
%
p_{r}(r,t)&=&-\frac{\dot{\Upsilon}(r,t)}{R^{2}\dot{R}}\label{eq10}\\
%
p_{t}(r,t)&=&-\frac{\dot{\Upsilon}(r,t)}{2R^{2}\dot{R}}-\frac{\Upsilon^{'}(r,t)}{2R\dot{R}\acute{R}}\label{eq11}
\end{eqnarray}

The field equation (\ref{eq4}) for radial pressure can be expressed in terms of area radius as 
\begin{equation}\label{eq12}
p_{r}=-\frac{1}{8\pi G}\left[2\frac{\ddot{R}}{R}+ \frac{\dot{R}^{2}}{R^{2}} + \frac{b(r)}{R^{2}}\right]
\end{equation}

which has a first integral as 
 \begin{equation}\label{eq13}
\dot{R}^{2}=-b(r)+\frac{c(r)}{R}-\frac{8\pi G}{R}\int p_{r}R^{2}dR
\end{equation}

with $c(r)$ an arbitrary integration function. 

As $p_{r}$ is regular initially at the center and blows up at the singularity so $p_{r}$ can be chosen as \cite{Chakraborty05} 
\begin{equation}\label{eq14}
p_{r}=\frac{p_{0}(r)}{R^{n}},
\end{equation}

 where $p_{o}$ is an arbitrary function of radial co-ordinate $r$ and $n$ is any constant. Note that $p_{o}\approx r^{n}$ near $r=0$ to make initial matter distribution to be non-zero at the center $r=0$.
 
As a result the evolution equation (\ref{eq13}) becomes  
\begin{equation}\label{eq15}
~~~~\dot{R}^{2}=-b(r)+\frac{c(r)}{R}-\frac{8\pi G}{R^{n-2}}\frac{p_{0}(r)}{(3-n)}, ~~~~~~~~~~~~~(n\neq 3)
\end{equation}

and the explicit form of $p_{t}$ reads
\vspace{-.3cm}
\begin{equation}\label{eq16}
p_{t}=\left(1-\frac{n}{2}\right)\frac{p_{0}(r)}{R^n}+\frac{r}{2}\frac{p_{0}^{'}(r)}{R^n}
\end{equation}

Now for smooth initial data $c(r)$, $p_{0}(r)$ and $b(r)$ to be $C^{\infty}$ functions and hence one has the Taylor series expansions
\vspace{-.2cm} 
\begin{eqnarray}\label{eq17}
 c(r)&=&\sum^{\infty}_{j=0}c_{j}r^{3+j}
\nonumber\\
 p_{0}(r)&=&\sum^{\infty}_{j=0}{p_{0}}_{j}r^{2+j}
\end{eqnarray}

\vspace{-.2cm}
and
\vspace{-.7cm}
\begin{eqnarray}
 b(r)&=&\sum^{\infty}_{j=0}b_{j}r^{2+j}
\nonumber
\end{eqnarray}

and hence \vspace{-.6cm}
\begin{eqnarray}\label{eq18}
\rho_{i}(r)&=&\sum^{\infty}_{j=0}\rho_{j}r^{j}\\
p_{t_{i}}(r)&=&\sum^{\infty}_{j=0}p_{Tj}r^{j}
\nonumber
\end{eqnarray}

Here the coefficients in the above expansions are all constants.

Initially choosing $R=r$, the matter density and the radial and tangential stresses have the initial values at the beginning of the collapsing process as
\begin{eqnarray}\label{eq19}
\rho_{i}(r)&=&\frac{{b(r)+rb^{'}(r)}}{r^{2}}
\nonumber\\
{p_{r}}_{i}(r)&=&\frac{p_{0}(r)}{r^{n}}
\\
{p_{t}}_{i}(r)&=&\left(1-\frac{n}{2}\right)\frac{p_{0}(r)}{r^{n}}+\frac{1}{2}\frac{p_{0}^{'}(r)}{r^{n-1}}
\nonumber
\end{eqnarray}

Now, the shell focusing singularity $t=t_{s}(r)$ is a hypersurface, characterized by 
 \begin{equation}\label{eq20}
R(t_{s}(r),r)=0
\end{equation}

and one gets the solution of the evolution equation (\ref{eq15}) in integral form as 
\begin{equation}\label{eq21}
t_{s}(r)-t_{i}=\int^{R}_{0}\frac{dR}{\left[-b(r)+\frac{c(r)}{R}-\frac{8\pi G}{R^{n-2}}\frac{p_{0}(r)}{3-n}\right]^{\frac{1}{2}}}
\end{equation}

where $t_{i}$ is the time of beginning of the collapse.

\section{Collapse Dynamics and Formation of Trapped Surface Formation}
\paragraph{  }In the collapse dynamics, event horizon for an observer at infinity has a crucial role to identify the nature of the singularity. Due to global nature of the event horizon and its formation depends very much on the construction of the null geodesics so for the present space-time geometry it is impossible to compute event horizon. Rather, we introduce a closely related concept- the notion of a trapped surface which is a compact space-like $2$-surface having normals on both sides are future pointing converging null geodesics families. Thus if the matter density falls off fast enough at infinity then a $2$ surface $S_{r,t}(r$=const, $t$=const) is a trapped surface having entire future development lie behind the event horizon. So, mathematically the normal null geodesics to the trapped surface $S_{r,t}$ is characterized by the tangent vector field $K^{\mu}$	satisfying $K_{\mu}K^{\nu}=0$ (null character), $K^{\mu}_{;\nu}K^{\nu}=0$ (geodesic) and $K^{\mu}_{;\mu}<0$ (convergent) \cite{Chakraborty05}.

It is to be noted that inward null geodesics converge initially and throughout the collapsing process while the outward geodesics diverge initially but become convergent after a time $t_{ah}(r)$, the time of formation of apparent horizon. The apparent horizon is characterized by \cite{Chakraborty05}
\begin{equation}\label{eq22}
\dot{R}(t_{ah}(r),r)=-\sqrt{1-b(r)}
\end{equation}

Using (\ref{eq22}) into the evolution equation (\ref{eq15}) area-radius at the apparent horizon is determined by the algebraic equation (choosing $8\pi G=1$, in the rest of the paper) 
\begin{equation}\label{eq23}
p_{0}(r)R^{3-n}\left(t_{ah}(r),r\right)+(n-3)c(r)-(n-3)R\left(t_{ah},r\right)=0
\end{equation}

As the integral equation (\ref{eq21}) or the algebraic equation (\ref{eq23}) cannot be solvable for any general `$n$' so we shall consider the following cases:

\textbf{Case-I: $n=2$}
 
In this case, the explicit solution for R by solving equation (\ref{eq15}) can be written as
\begin{equation}\label{eq24a}
t(r)-t_{i}=\frac{c(r)}{\mu^{\frac{3}{2}}(r)}\left[\sin^{-1}\sqrt{\frac{\mu(r)r}{c(r)}}-\sqrt{\frac{\mu(r)r}{c(r)}}\sqrt{1-\frac{\mu(r)r}{c(r)}}\right]-\frac{c(r)}{\mu^{\frac{3}{2}}(r)}\left[\sin^{-1}\sqrt{\frac{\mu(r)R}{c(r)}}-\sqrt{\frac{\mu(r)R}{c(r)}}\sqrt{1-\frac{\mu(r)R}{c(r)}}\right]
\end{equation}

The coefficients of the series expansion are related among themselves by the relations (\ref{eq17}) and (\ref{eq18}) as follows:-
\begin{eqnarray}\label{eq24}
\rho_{0}&=&3b_{0},~~~~~~~\rho_{1}=4b_{1},~~~~~~\rho_{2}=5b_{2}
\nonumber\\
p_{T0}&=&p_{00},~~~~~p_{T1}=\frac{3}{2}p_{01},~~~~~p_{T2}=2p_{02}\\
\mu_{0}&=&\frac{\rho_{0}}{3}+p_{T0},~~~~~~\mu_{1}=\frac{\rho_{1}}{4}+\frac{2p_{T1}}{3}
\nonumber
\end{eqnarray}

In this case the singularity hypersurface can be written in explicit form as 
\begin{equation}
t_{s}(r)-t_{i}=\frac{c(r)}{\mu^{\frac{3}{2}}(r)}\left[\sin^{-1}\sqrt{\frac{\mu(r)r}{c(r)}}-\sqrt{\frac{\mu(r)r}{c(r)}}\sqrt{1-\frac{\mu(r)r}{c(r)}}~\right]
\end{equation}

where $\mu(r)=p_{0}(r)+b(r).$

Hence using equations (\ref{eq22}) and (\ref{eq23}) the time of formation of trapped surface is given by
\begin{equation}\label{eq26}
t_{ah}(r)-t_{i}=\frac{c(r)}{\mu^{\frac{3}{2}}(r)}\left[\sin^{-1}\sqrt{\frac{\mu(r)r}{c(r)}}-\sqrt{\frac{\mu(r)r}{c(r)}}\sqrt{1-\frac{\mu(r)r}{c(r)}}\right]-\frac{c(r)}{\mu^{\frac{3}{2}}(r)}\left[\sin^{-1}\sqrt{\mu(r)}-\sqrt{\mu(r)}\sqrt{1-\mu(r)}\right]
\end{equation}

The  time of occurrence of central cell focusing singularity (at $r=0$) will be given by 
\begin{equation}\label{eq27}
t_{0}=\lim_{r\rightarrow 0} t_{s}(r) = t_{i}+ \frac{c_{0}}{\mu_{0}^{\frac{3}{2}}}\left[\sin^{-1}\sqrt{x_{0}}-\sqrt{x_{0}}\sqrt{1-x_{0}}\right]
\end{equation}

where $x_{0}=\dfrac{\mu_{0}}{c_{0}}$
and in evaluating the limit we have used the series form of $\mu(r)$, $b(r)$ and $c(r)$ from equation(17).

Thus the time difference between $t_{ah}(r)$ and $t_{o}$ is given by
\begin{equation}
t_{ah}(r)-t_{0}=\frac{c_{0}}{\mu_{0}}\left[\overline{A}+\frac{\mu_{0}}{2}\frac{1-2x_{0}}{\sqrt{1-x_{0}}}\left(\frac{c_{1}}{c_{0}}-\frac{\mu_{1}}{\mu_{0}}\right)+\left(\frac{c_{1}}{c_{0}}-\frac{3\mu_{1}}{2\mu_{0}}\right)\left[\sin^{-1}\sqrt{x_{0}}-\sqrt{x_{0}(1-x_{0})}\right]\right]r+\mathcal{O}(r^{2})
\end{equation}
where $\overline{A}=
\dfrac{c_{0}}{2}\left(\dfrac{\mu_{1}}{\mu_{0}}-\dfrac{c_{1}}{c_{0}}\right)\left[\sqrt{x_0}(1+x_0)\cosh(\sqrt{x_0})-x_0\sinh(\sqrt{x_0})\right]$ and $x_{0}=\dfrac{\mu_{0}}{c_{0}}$.

It is to be noted that $t_{0}$ is the time of formation of singularity at $r=0$ while $t_{ah}(r)$ is the epoch at which a trapped surface is formed at a distance $r$.Thus if trapped surface is formed at a later instant than $t_{0}$ then it is possible that any light signal from the singularity can reach an observer. Therefore $t_{ah}(r)>t_{0}$ is the necessary condition for formation of naked singularity, while to form black hole, the sufficient condition is $t_{ah}(r)\leq t_{0}$. It should be mentioned that this criterion for naked singularity is purely local.

For $x_{0}\ll 1$ , the time difference can be approximately written as 
$$t_{ah}-t_{0}\approx\frac{c_{0}}{\mu_{0}^{\frac{3}{2}}}\frac{\mu_{0}}{2}\left[\frac{c_{1}}{c_{0}}-\frac{\mu_{1}}{\mu_{0}}\right]r$$

Hence for naked singularity i.e. $t_{ah}>t_{0}$ one must have
 $$\frac{c_{1}}{c_{0}}>\frac{\frac{\rho_{1}}{4}+\frac{2p_{T1}}{3}}{\frac{\rho_{0}}{3}+p_{T0}}$$

The above relation indicates that formation of black hole or naked singularity depends on the interrelation among the coefficients of arbitrary integration constant, density and radial pressure.

\textbf{Case:-II  $n=4$}

In this case, using evolution equation (\ref{eq15}) the explicit solution for R can be written as
\begin{equation}
t(r)-t_{i}=\frac{1}{b(r)}\left[\sqrt{A-\left({R}-\frac{c}{2b}\right)^{2}}-\sqrt{A-\left(r-\frac{c}{2b}\right)^{2}}\right]+\frac{c(r)}{2b^{\frac{3}{2}}}\left[\sin^{-1}\frac{r-\frac{c}{2b}}{\sqrt{A}}-\sin^{-1}\frac{R-\frac{c}{2b}}{\sqrt{A}}\right]
\end{equation}

 The coefficients of the series expansion are related among themselves through the relations (\ref{eq17}) and  (\ref{eq18}) as follows:-  \vspace{-.2cm}                           
\begin{eqnarray}\label{eq29} 
\rho_{0}&=&3b_{0},~~~~~~~\rho_{1}=4b_{1},~~~~~~\rho_{2}=5b_{2}
\nonumber\\
p_{T0}&=&p_{00},~~~~~~p_{T1}=4p_{01},~~~~~p_{T2}=5p_{02}\\
\mu_{0}&=&\frac{\rho_{0}}{3}+p_{T0},~~~~~~\mu_{1}=\frac{\rho_{1}}{4}+\frac{p_{T1}}{4} 
\nonumber
\end{eqnarray}

In this case, the singularity hypersurface can be written in explicit form as 
\begin{eqnarray}\label{eq30}
t_{s}(r)-t_{i}=\left[\frac{1}{2b_{0}^{\frac{3}{2}}}c_{0}\left(\sin^{-1}\mu_{0}+\sin^{-1}\lambda_{0}\right)+B_{0}D_{0}\right]+\Bigg[\frac{1}{2b_{0}^{\frac{3}{2}}}\bigg\{c_{0}\left(A_\mu+A_{\lambda}\right)+\left(\sin^{-1}\mu_{0}+\sin^{-1}\lambda_{0}\right)
\nonumber
\\
\left(c_{1}-\frac{3b_{1}c_{0}}{2b_{0}}\right)\bigg\}+(D_{1}-B_{1})\Bigg]r+\mathcal{O}\left(r^{2}\right)~~ 
\end{eqnarray}


For detail expression for each variable see Appendix \ref{A}.

The time of formation of trapped surface is given by
\begin{equation}\label{eq31}
t_{ah}(r)-t_{i}=\frac{1}{b(r)}\left[\sqrt{A-\left(R_{ah}-\frac{c}{2b}\right)^{2}}-\sqrt{A-\left(r-\frac{c}{2b}\right)^{2}}\right]+\frac{c(r)}{2b^{\frac{3}{2}}}\left[\sin^{-1}\frac{r-\frac{c}{2b}}{\sqrt{A}}-\sin^{-1}\frac{R_{ah}-\frac{c}{2b}}{\sqrt{A}}\right]
\end{equation}

where $A=\frac{p_{0}(r)}{b(r)}+\frac{c^{2}(r)}{4b^{2}(r)}$.

The time of occurrence of the central shell focusing singularity (at $r=0$) will be given by
\begin{equation}\label{eq32}
t_{0}=\lim_{r\rightarrow 0}t_{s}(r)= t_{i}+ \frac{c_{0}}{2b_{0}^{\frac{3}{2}}}(sin^{-1}\mu_{0}+sin^{-1}\lambda_{0})+B_{0}D_{0}
\end{equation}

Now the time difference between the formation of trapped surface and central shell focusing singularity is given by
\begin{eqnarray}\label{eq33}
t_{ah}(r)-t_{o}=\frac{1}{\sqrt{b_{o}}}\left[\sqrt{A_{o}-\frac{c_{o}^{2}}{4b_{o}^{2}}}+\left(A_{5}-\frac{b_{1}}{2b_{o}}\sqrt{A_{o}-\frac{c_{o}^{2}}{4b_{o}^{2}}}\right)r\right]-B_{o}-B_{1}r+\frac{1}{2b_{o}^{\frac{3}{2}}}\left(c_{o}+\left(c_{1}-\frac{3b_{1}c_{o}}{2b_{o}}\right)r\right)
\nonumber\\
\left[\sin^{-1}\mu_{0}-\sin^{-1}\xi_{0}+\left(A_{\mu}-A_{\xi}\right)r\right]-\frac{c_{0}}{2b_{0}^{\frac{3}{2}}}\left[\sin^{-1}\mu_{0}+\sin^{-1}\lambda_{0}\right]-B_{0}D_{0}~~
\end{eqnarray}

where \vspace{-.5cm}$$A_{\xi}=
\xi_{1}\left[(1+\xi_0^2)\cosh(\xi_0)-\xi_0\sinh(\xi_0)\right]$$
$$A_{5}=\frac{A_{1}+\frac{c_{0}}{\sqrt{b_{0}}}\left(\sqrt{p_{00}}-\frac{c_{1}}{2b_{0}}+\frac{c_{0}b_{1}}{2b_{0}^{2}}\right)}{2\left(A_{0}-\frac{c_{0}^{2}}{4b_{0}^{2}}\right)}~~~~$$
$$\xi_{0}=-\frac{c_{0}}{2b_{0}\sqrt{A_{0}}}~~~~~~~~~~~~~~~~~~~~~~~~~~~~~$$
$$~~~~\xi_{1}=\frac{1}{\sqrt{A_{0}}}\left[\sqrt{p_{00}}-\frac{c_{1}}{2b_{0}}+\frac{c_{0}b_{1}}{2b_{0}^{2}}+\frac{c_{0}A_{1}}{4b_{0}A_{0}}\right]$$

Due to the complicated form of the above equation, it is very difficult to make a comparative study between $t_{ah}$ and $t_{0}$. Figure \ref{f1} shows the time difference graphically for some choices of the parameter and it favored mostly for the formation of black hole.
	\begin{figure}[h]
	\begin{center}
		\includegraphics[height=.25\textheight,width=.35\textheight]{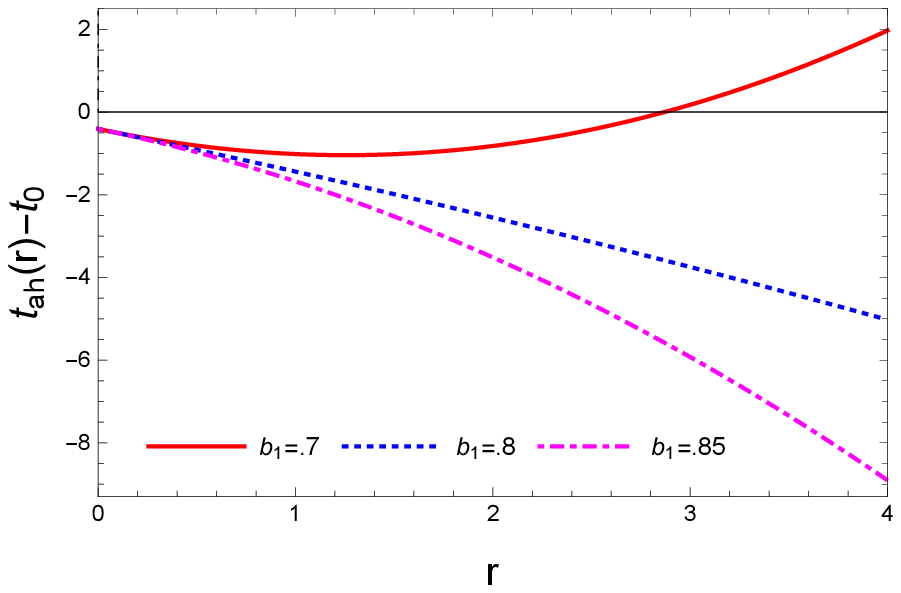}
	\end{center}
	\vspace{-.5cm}\hspace{2cm}
	\begin{minipage}{.7\textwidth}
		\caption{\small\textsf{The figure shows the time difference between the formation of trapped surface and central shell focusing singularity given in equation (\ref{eq33}) for the choice of the parameter $ p_{00}=.4,~ c_0=.35	,~ b_0=0.5,~c_1=-.7	,~	 p_{01}= 1.$}}
		\label{f1}
		\end{minipage}
\end{figure}

\textbf{Case:-III}  For General `$n$'

The evolution equation (\ref{eq15}) for $R$ can only be solvable for general values of `n', provided the arbitrary integration function c(r)=0. The solution takes the form
\begin{equation}\label{eq34}
t(r)-t_{i}=\frac{R(t,r)}{\sqrt{b(r)}}2F_{1}\left[\frac{1}{2},\frac{1}{2-n},1+\frac{1}{2-n},\frac{R^{2-n}p_{0}(r)}{(3-n)b(r)}\right]- \frac{r}{\sqrt{b(r)}}2F_{1}\left[\frac{1}{2},\frac{1}{2-n},1+\frac{1}{2-n},\frac{r^{2-n}p_{0}(r)}{(3-n)b(r)}\right]
\end{equation}
with $b(r)<0$,  $n<3$ and $n\neq~2$.

Hence the singularity hypersurface can be written in explicit form as 
\begin{equation}\label{eq35}
t_{s}(r)-t_{i}=-\frac{r}{\sqrt{b(r)}}2F_{1}\left[\frac{1}{2},\frac{1}{2-n},1+\frac{1}{2-n},\frac{r^{2-n}p_{0}(r)}{(3-n)b(r)}\right]
\end{equation}

Now, using evolution equation (\ref{eq15}) the time of formation of apparent horizon has the explicit form 
\begin{equation}\label{eq36}
t_{ah}(r)-t_{i} =\frac{r}{\sqrt{b(r)}}2F_{1}\left[\frac{1}{2},\frac{1}{2-n},1+\frac{1}{2-n},\frac{r^{2-n}p_{0}(r)}{(3-n)b(r)}\right]-\frac{R(t_{ah},r)}{\sqrt{b(r)}}2F_{1}\left[\frac{1}{2},\frac{1}{2-n},1+\frac{1}{2-n},\frac{R^{2-n}(t_{ah},r)p_{0}(r)}{(3-n)b(r)}\right]
\end{equation}

The central shell focusing singularity (at $r=0$) will occur at time $t_{o}$ given by 
\begin{equation}\label{eq37}
t_{0}=\lim_{r\rightarrow0}t_{s}(r)=t_{i}+\frac{1}{\sqrt{b_{0}}}2F_{1}\left[\frac{1}{2},\frac{1}{2-n},1+\frac{1}{2-n},\frac{p_{00}b_{0}}{3-n}\right]
\end{equation}

where the series expansions $p_0(r)=\sum\limits^\infty_{j=0}p_{0j}r^{n+j}$,~~~$b(r)=\sum\limits^\infty_{j=0}b_{j}r^{2+j}$ have been used to evaluate the limit. Hence the time difference between the formation of apparent horizon and the central singularity is given by
\begin{eqnarray}\label{eq38}
t_{ah}(r)-t_{0}=\Bigg[b_{0}^{-\frac{5}{2}}2F_{1}\left[\frac{3}{2},1+\frac{1}{2-n},2+\frac{1}{2-n},\frac{p_{00}}{(3-n)b_{0}}\right]\left(p_{01}b_{0}-p_{00}b_{1}\right)-\frac{b_{0}^{-\frac{3}{2}}b_{1}}{2}~~~~~~~~~~~~~~~~~~~~~~~~~~~~~~~~~~~~~~~
\nonumber\\
2F_{1}\left[\frac{1}{2},\frac{1}{2-n},1+\frac{1}{2-n},\frac{p_{00}}{(3-n)b_{0}}\right]\Bigg]r+\mathcal{O}(r^{2})-\frac{R_{0}}{\sqrt{b_{0}}}r^{\frac{n}{n-2}-3}2F_{1}\left[\frac{1}{2},\frac{1}{2-n},1+\frac{1}{2-n},\frac{p_{00}}{(3-n)b_{0}}\right]
\end{eqnarray}

Although the above time difference is very complicated in form, so to make some conclusion we have presented the time difference graphically in Figure \ref{f2}. In this case also, the formation of black hole is most dominating than the formation of naked singularity.
	\begin{figure}[h]
			\begin{minipage}{0.45\textwidth}
		\includegraphics[height=.25\textheight,width=.35\textheight]{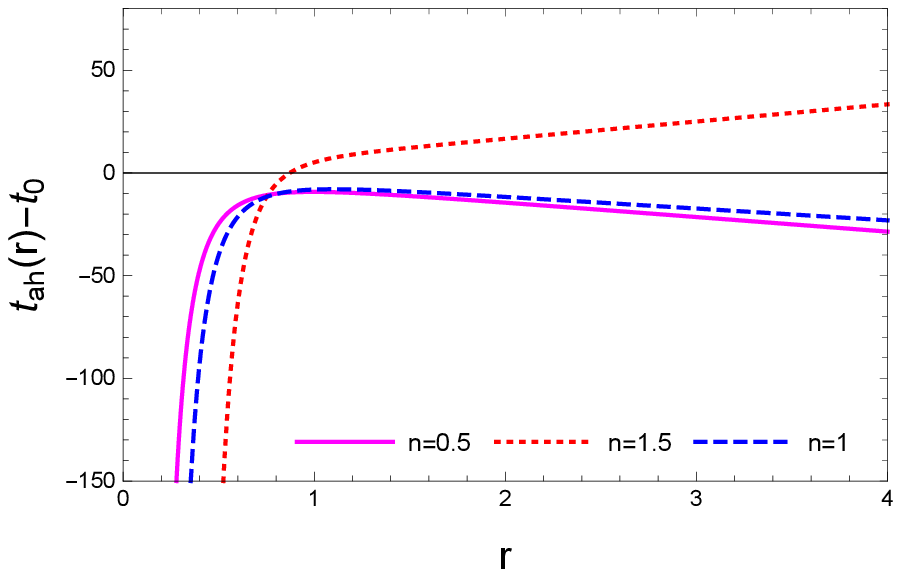}
\end{minipage}~~~~~~~~~~
		\begin{minipage}{0.45\textwidth}
	\includegraphics[height=.25\textheight,width=.35\textheight]{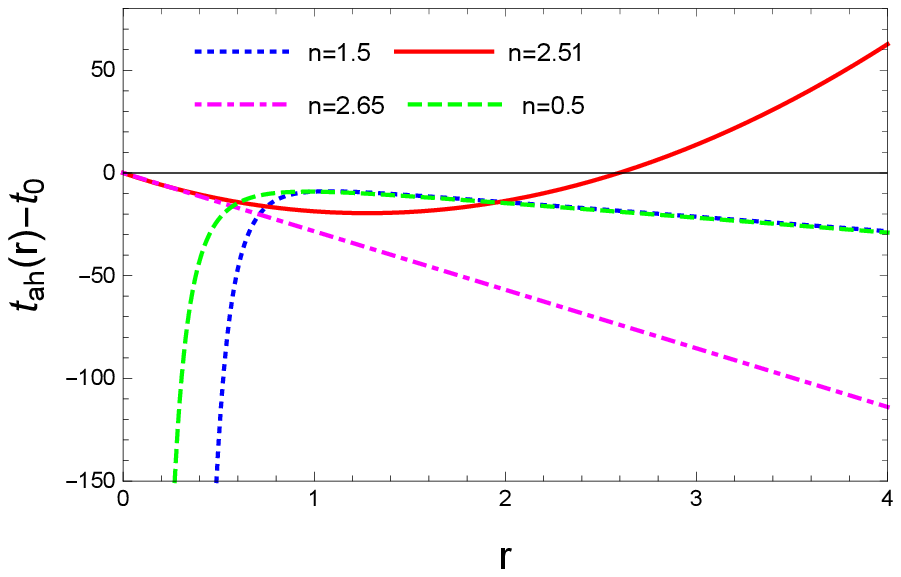}
\end{minipage}
\vspace{-.5cm}\hspace{1.3cm}
\begin{minipage}{.8\textwidth}
	\caption{\small\textsf{The figure shows the time difference between the formation of trapped surface and central shell focusing singularity given in equation (\ref{eq38}) for the choice of the parameter $b_0 = 0.3,~	 b_1= 0.2,	 ~p_{01}= 0.85,~	 R_0 = 1,~p_{00}=0 .4$(left panel),$~p_{00}=0 .1$(right panel)}}
	\label{f2}
\end{minipage}
\end{figure}
\vspace{-.5cm}
\section{Geodesics near Singularity}
\paragraph{  }    This section will discuss the nature of singularity (naked or covered) by examining the null geodesics. In particular, it will be investigated whether there exists outgoing radial null geodesics (ORNG) which terminate in the past at the central singularity. Thus assuming the existence of ORNG through central singularity: $r=0$ in the past one has (near $r=0$) \cite{Chakraborty05}
\begin{equation}\label{eq39}
	t_{ORNG}=t_{0}+\lambda~r^{\alpha}
\end{equation}

up to leading order in (r,t)-plane with $\lambda>0$, $\alpha>0$. As the evolution equation of $R$ (i.e. eq (\ref{eq15})) cannot be solvable in general so to have an analytic solution we consider the following cases:
a) $n=2$, b) $n=4$

\textbf{Case(a): $n=2$}

The solution to the evolution equation (\ref{eq15}) can be written as (choosing $t_{i}=0$) 
\begin{equation}\label{eq40}
t(r)=\frac{c}{\mu^{\frac{3}{2}}}\left[\left[\frac{\sqrt{\mu~R(c-\mu~R)}}{c}-\sin^{-1}\sqrt{\frac{\mu~R}{c}}\right]-\left[\frac{\sqrt{\mu~r(c-\mu~r)}}{c}-\sin^{-1}\sqrt{\frac{\mu~r}{c}}\right]\right]
\end{equation}.

So the expression for singularity time (defined as $R(t_{s}(r),r)=0$) is given by 
\begin{equation}\label{eq41}
t_{s}(r)=\frac{c}{\mu^{\frac{3}{2}}}\left[-\frac{\sqrt{\mu~r(c-\mu~r)}}{c}+\sin^{-1}\sqrt{\frac{\mu~r}{c}}\right]
\end{equation}

Hence the time for central singularity is 
\begin{equation}\label{eq42}
t_{0}=\lim_{r\rightarrow0}t_{s}(r)=\frac{c_{0}}{\mu_{0}^{\frac{3}{2}}}\left[\sin^{-1}\sqrt{\frac{\mu_{0}}{c_{0}}}-\sqrt{\frac{\mu_{0}}{c_{0}}\left(1-\frac{\mu_{0}}{c_{0}}\right)}\right]
\end{equation}

In deriving the above limiting values one has to use the following finite series forms
\begin{equation}\label{eq43}
c(r)=c_{o}r^{3}+c_{k}r^{3+k},~  \mu(r)=\mu_{o}r^{2}+\mu_{l}r^{l+2}
\end{equation}

Here all the coefficients are constants with $c_{k}(<0)$ and $\mu_{l}(<0)$, the first non-vanishing term beyond $c_{o}$ and $\mu_{o}$ respectively.
Now using (\ref{eq43}) in (\ref{eq41}) the singularity time takes the form 
\begin{equation}\label{eq44}
t_{s}(r)=t_{0}+D_{l}r^{l}+F_{k}r^{k}
\end{equation}

with \vspace{-.5cm}
\begin{eqnarray}\label{eq45}
D_{l}&=&\frac{c_{0}}{\mu_{0}^{\frac{3}{2}}}\left[A_{l}-B_{l}-\frac{3}{2}\frac{\mu_{l}}{\mu_{0}}sin^{-1}\sqrt{x_{0}}\right]
\nonumber\\
F_{k}&=&\frac{1}{\mu_{0}^{\frac{3}{2}}}\left[c_{0}(A_{k}-B_{k})+c_{k}sin^{-1}\sqrt{x_{0}}\right]
\end{eqnarray}

where \vspace{-.5cm}$$~~~~~~~~~~A_{l}=
\frac{x_l}{2}\left[\left(\sqrt{x_0}+\frac{1}{\sqrt{x_0}}\right)\cosh(\sqrt{x_0})-\sinh(\sqrt{x_0})\right]$$
$$B_{l}=\frac{x_{l}}{2}\frac{1-2x_{0}}{x_{0}(1-x_{0})},~ x_{0}=\frac{\mu_{0}}{c_{0}}, ~x_{l}=-\frac{\mu_{0}c_{l}}{c_{0}^{2}}$$

So there are two possibilities namely (i) $K<l$  and (ii) $K>l$.
For $K<l$ , one has 
\begin{equation}\label{eq46}
t_{s}(r)=t_{0}+F_{k}r^{k},
\end{equation}

which in comparison with the geodesic equation (\ref{eq39}) gives $\alpha~>K$ or $\alpha~=K$ and $\lambda~<B_{0}$. 
Thus for $\alpha~>K$ the solution for $R$ near $r=0$ simplifies to
\begin{eqnarray}\label{eq47}
R&=&3\left[\left(\frac{c}{\mu}\right)^{\frac{3}{2}}sin^{-1}\sqrt{\frac{\mu~r}{c}}-\left(\frac{c}{\mu}\right)^{\frac{3}{2}}\sqrt{\frac{\mu~r}{c}\left(1-\frac{\mu~r}{c}\right)}-\sqrt{c}t\right]^{\frac{2}{3}}
\nonumber\\
&\approx&r\left[3\left[\left(\frac{c_{0}}{\mu_{0}}\right)^ {\frac{3}{2}}sin^{-1}\sqrt{\frac{\mu_{0}}{c_{0}}}-\left(\frac{c_{0}}{\mu_{0}}\right)^{\frac{3}{2}}\sqrt{\frac{\mu_{0}}{c_{0}}\left(1-\frac{\mu_{0}}{c_{0}}\right)}-\sqrt{c_{0}}t\right]^{\frac{2}{3}}\right]
\end{eqnarray}

Hence it is possible to have radial null geodesic in this case.

The other case for $K>l$ is similar to above.

\textbf{Case(b): $n=4$}

The solution to the evolution equation (\ref{eq15}) can be written as (choosing $t_{i}=0$) 
\begin{equation}\label{eq48}
t(r)=\frac{1}{\sqrt{b(r)}}\sqrt{A-\left(R-\frac{c(r)}{2b(r)}\right)^{2}}-\frac{1}{\sqrt{b(r)}}\sqrt{A-\left(r-\frac{c(r)}{2b(r)}\right)^{2}}+\frac{c}{2b^{3/2}}\left[\frac{\sin^{-1}\left(r-\frac{c}{2b}\right)}{\sqrt{A}}-\frac{\sin^{-1}\left(R-\frac{c}{2b}\right)}{\sqrt{A}}\right]
\end{equation}

where $A=\frac{p_{0}(r)}{b(r)}+\frac{c^{2}(r)}{4b^{2}(r)}$.

So the expression for singularity time (defined as $R(t_{s}(r),r)=0$) is given by 
\begin{equation}\label{eq49}
t_{s}(r)-t_{i}=\frac{c}{2b^{\frac{3}{2}}}\left[\frac{\sin^{-1}(r-\frac{c}{2b})}{\sqrt{A}}+\frac{\sin^{-1}c}{2b\sqrt{A}}\right]+\frac{\sqrt{A-\left(\frac{c}{2b}\right)^{2}}}{\sqrt{b}}-\frac{\sqrt{A-\left(r-\frac{c}{2b}\right)^{2}}}{b}
\end{equation}

Hence the time for central singularity is 
\begin{equation}\label{eq50}
t_{0}=\lim_{r\rightarrow0}t_{s}(r)=t_{i}+\frac{c_{0}}{2b_{0}^{\frac{3}{2}}}\left[\sin^{-1}\overline{\mu_{0}}+\sin^{-1}\overline{\lambda_{0}}\right]+\frac{\sqrt{p_{4}B_{0}}}{b_{0}}
\end{equation}

In deriving the above limiting values one has to use the following finite series forms
\begin{equation}\label{eq51}
c(r)=c_{0}r^{3}+c_{k}r^{3+k},~  p_{0}(r)=p_{4}r^{4}+p_{m}r^{4+m},  ~   b(r)=b_{0}r^{2}+b_{l}r^{l+2}
\end{equation}

Here all the coefficients are constants with $c_{k}(<0)$, $p_{4}(<0)$ and $b_{l}(<0)$, the first non-vanishing terms beyond $c_{0}$, $p_{0}$ and $b_{0}$ respectively.

Now using (\ref{eq51}) in (\ref{eq49}) the singularity time takes the form 
\begin{eqnarray}\label{eq52}
t_{s}(r)=t_{0}+r^{k}\left[c_{k}\left(sin^{-1}\mu_{0}+sin^{-1}\lambda_{0}\right)+\frac{\overline{B_{2}}\sqrt{p_{4}}}{b_{0}}\right]-\frac{3}{4}\frac{b_{l}r^{l}}{b^{\frac{5}{2}}_{0}}r^{l}\left(\sin^{-1}\mu_{0}+\sin^{-1}\lambda_{0}\right)+
\nonumber\\
\left[\frac{c_{0}}{2b_{0}^{\frac{3}{2}}}\left(A_{\overline{\mu_{3}}}+A_{\overline{\lambda_{3}}}\right)+\left(\frac{B_{0}p_{m}}{2\sqrt{p_{4}}b_{0}}+\frac{\overline{B_{3}}\sqrt{p_{4}}}{b_{0}}\right)\right]r^{m}+\left(\frac{\overline{B_{1}}\sqrt{p_{4}}}{b_{0}}-\frac{2b_{l}p_{4}^{\frac{3}{2}}}{b_{0}}\right)r^{l}
\end{eqnarray}

For detail expression for each variable see Appendix \ref{B}

So, it is possible to have radial null geodesic near the central singularity for $n=4$.
\section{Summary and Remarks}
\paragraph{  }In the present work the inhomogeneous FRW model of the space-time has been considered for studying collapse dynamics. The basic question that one has to address in these studies is whether the end state of collapse will be a naked singularity or black hole will form to prevent the visualization of the singularity. It is not desirable from physical point of view that the singularity is naked and Penrose has termed this unphysical phenomenon as cosmic censorship conjecture (CCC). However, there are various examples in the literature which contradict this CCC. In this study, the validity of this conjecture has been examined either by calculating the time difference between the formation of central singularity and the formation of apparent horizon (positivity of this time difference implies naked singularity while negativity indicates formation of black hole) or by studying the feasibility of constructing null geodesics originated from the central singularity. In particular, naked singularity or black hole formation depends on the initial matter distribution which in the present case is chosen as anisotropic perfect fluid.
   
 Due to complicated form of the field equations the evolution equation for area radius $R$ cannot be solved in general. However, for two particular choice of the parameter involved one has an explicit analytic solution for $R$. But the analytic expression of the time difference between the formation of singularity and apparent horizon is also very complicated and one cannot have any definite conclusion for the end state of collapse. However graphically we have plotted the time difference for some suitable choice of parameters involved and it is found that our model is favorable for the formation of the black hole rather than naked singularity. Similar is the situation in examining the possible geodesic through the resulting singularity. From both these aspects there is no definite interference about the end phase of the collapsing process and also it is not possible to identify the definite role of the matter distribution on the final state of the collapsing process. Finally, one cannot definitely conclude in favor or against the cosmic censorship conjecture due to Penrose.
 
}

{\Large\textbf{Appendix}}
\appendix
\section{Detail expressions for each variable in equation (\ref{eq30})}\label{A}
 $$A_{0}=\frac{p_{00}}{b_{0}}+\frac{c_{0}^{2}}{4b_{0}^{2}},~~~~~~~~~~~~~~~~~~~~~~~~~~~~~~~~~~~~~
A_{1}=\left(\frac{p_{01}}{b_{0}}-\frac{p_{00}b_{1}}{b_{0}^{2}}\right)+\frac{c_{0}}{2b_{0}}\left(\frac{c_{1}}{b_{0}}-\frac{c_{0}b_{1}}{b_{0}^{2}}\right),$$
$$A_{\mu}=
\mu_{1}\left[(1+\mu_0^2)\cosh(\mu_0)-\mu_0\sinh(\mu_0)\right],~~~
A_{\lambda}
=\lambda_{1}\left[(1+\lambda_0^2)\cosh(\lambda_0)-\lambda_0\sinh(\lambda_0)\right],$$
$$B_{0}=\sqrt{\frac{A_{0}-1+\frac{c_{0}}{2b_{0}}}{b_{0}}},~~~~~~~~~~~~~~~~~~~~~~~~~~~~~~
B_{1}=\frac{B_{0}}{2}\frac{A_{1}+\frac{c_{1}}{2b_{0}}-\frac{b_{1}A_{0}}{b_{0}}+\frac{b_{1}}{b_{0}}-\frac{c_{0}b_{1}}{2b_{0}^{2}}}{A_{0}-1+\frac{c_{0}}{2b_{0}}},~~~~~$$
$$\lambda_{0}=\frac{c_{0}}{2b_{0}\sqrt{A_{0}}},~~~~~~~~~~~~~~~~~~~~~~~~~~~~~~~~~~~~~~~~
\lambda_{1}=\frac{\left[c_{1}-c_{0}\left(\frac{b_{1}}{b_{0}}+\frac{A_{1}}{2A_{0}}\right)\right]}{2b_{0}\sqrt{A_{0}}},~~~~~~~~~~~~~~~~~~~$$
$$~~~~\mu_{0}=\frac{\left(1-\frac{c_{0}}{2b_{0}}\right)}{\sqrt{A_{0}}},~~~~~~~~~~~~~~~~~~~~~~~~~~~~~~~~
~~~~~\mu_{1}=\frac{1}{2\sqrt{A_{0}}}\left[\left(\frac{c_{0}b_{1}}{b_{0}^{2}}-\frac{c_{1}}{b_{0}}\right)+\frac{c_{0}}{2b_{0}A_{0}}-\frac{A_{1}}{A_{0}}\right],~~$$
$$D_{0}=\frac{\sqrt{p_{00}}}{b_{0}},~~~~~~~~~~~~~~~~~~~~~~~~~~~~~~~~~~~~~~~~~~~
D_{1}=\frac{1}{2\sqrt{p_{00}}}\left(\frac{p_{01}}{b_{0}}-\frac{p_{00}b_{1}}{b_{0}^{2}}-\frac{b_{1}}{b_{0}}\right),~~~~~~~~~~~$$
\section{Detail expressions for each variable in equation (\ref{eq52})}\label{B}
$$~~~~~~A_{0}=\frac{p_{0}}{b_{0}}+\frac{c_{0}^{2}}{4b_{0}^{2}},~~~~~~~~~~~~~~~~~~~~~~~~~~
B_{0}=\sqrt{\frac{A_{0}-1+\frac{c_{0}}{2b_{0}}}{b_{0}}}$$
$$~~~~~\overline{A_{1}}=-\left[\frac{p_{4}b_{l}}{b_{0}^{2}}+\frac{2b_{l}c_{0}^{2}}{4b_{0}^{3}}\right],~~~~~~~~~~~~~~~
\overline{A_{2}}=\frac{p_{m}}{b_{0}^{2}}~~~~~~~~~~~~~~$$
$$~~~~~~~~~~~~~~~~~~~~~~~~~~~\overline{A_{3}}=\frac{c_{k}c_{0}}{2b_{0}^{2}},~~~~~~~~~~~~~~~~~~
~~~~~~~~~~~~~~\overline{B_{1}}=\frac{\overline{B_{0}}}{2}\frac{\overline{A_{1}}-\frac{b_{l}c_{0}}{2b_{0}^{2}}-\frac{b_{l}A_{0}}{b_{0}}+\frac{b_{l}}{b_{0}}-\frac{b_{l}c_{0}}{2b_{0}^{2}}}{A_{0}-1+\frac{c_{0}}{2b_{0}}}$$
$$~~~~~~~~~~~\overline{B_{2}}=\frac{\overline{A_{2}}B_{0}}{2\left(A_{0}-1+\frac{c_{0}}{2b_{0}}\right)},~~~~~~~~~~~~~~~
\overline{B_{3}}=\frac{\left(\overline{A_{3}}+\frac{c_{k}}{2b_{0}}B_{0}\right)}{2\left(A_{0}-1+\frac{c_{0}}{2b_{0}}\right)}~~~$$
$$~~~~A_{\overline{\mu_{3}}}=-\frac{\overline{A_{2}}}{2A_{0}}\left(1-\frac{c_{0}}{2b_{0}}\right),~~~~~~~~~~~~~~
A_{\overline{\lambda_{3}}}=-\frac{c_{0}\overline{A_{2}}}{2A_{0}}~~~~~~~~~$$
\end{document}